# Electrochemical Impedance Imaging via the Distribution of Diffusion Times

Juhyun Song[1], and Martin Z. Bazant[1,2,*]

[1]*Department of Chemical Engineering* and [2]*Department of Mathematics,*

*Massachusetts Institute of Technology, Cambridge, Massachusetts 02139, United States*

ABSTRACT

We develop a mathematical framework to analyze electrochemical impedance spectra in terms of a distribution of diffusion times (DDT) for a parallel array of random finite-length Warburg (diffusion) or Gerischer (reaction-diffusion) circuit elements. A robust DDT inversion method is presented based on Complex Nonlinear Least Squares (CNLS) regression with Tikhonov regularization and illustrated for three cases of nanostructured electrodes for energy conversion: (i) a carbon nanotube supercapacitor, (ii) a silicon nanowire Li-ion battery, and (iii) a porous-carbon vanadium flow battery. The results demonstrate the feasibility of non-destructive "impedance imaging" to infer microstructural statistics of random, heterogeneous materials.

MAIN TEXT

Impedance spectroscopy is powerful method of non-destructive evaluation for electrochemical systems and materials, which relies on physics-based circuit models to interpret experimental data[1,2]. In pursuit of a higher power and energy density as well as a longer lifetime, electrochemical energy systems increasingly employ hierarchical nanostructured materials[3-6]: battery electrodes consist of nanoparticles[7-12]; supercapacitor electrodes are full of nanopores[13-17]; and in fuel cells[18-21] and flow batteries[22-24] electrolytes flow through nanoporous electrodes. The nanoscale diffusion lengths in such materials renders the low-frequency transition from infinite-length Warburg impedance, scaling as $(i\tau\omega)^{-1/2}$ where $\tau$ is the characteristic diffusion time and $\omega$ is the applied frequency, to finite-length behavior, either resistive or capacitive, now fully accessible to electrochemical impedance spectroscopy [25-27]. Conventional finite-length diffusion circuit elements are available to capture the transition, scaling as $\tanh(\sqrt{i\tau\omega})/\sqrt{i\tau\omega}$ for transmissive diffusion[28-30] and $\coth(\sqrt{i\tau\omega})/\sqrt{i\tau\omega}$ for bounded diffusion[26,31], and variants have been derived for nonplanar geometries, such as cylindrical, spherical, and rectangular shapes[27,32]. When a charge transfer reaction takes place simultaneously along with diffusion, their Gerischer-type derivatives are obtained by replacing $\sqrt{i\tau\omega}$ with $\sqrt{\tau(k+i\omega)}$, where $k$ is the apparent first order kinetic constant[25,33-35]. Identical expressions are also used to model the impedance of porous electrodes, where $\tau$ becomes the charging time of the RC transmission line representing the pore resistors and double layer capacitors [36,37].

Since conventional models assume constant material properties and simple geometries, experimental Nyquist plots often exhibit large deviations, such as asymmetric, depressed Warburg arc for transmissive diffusion[38-40] or an inclined capacitive rays for bounded diffusion[41-47]. The most common heuristic approach to describe such deviations is the constant phase element (CPE), $(i\tau\omega)^{-\alpha}$ where $0<\alpha<1$, placed in various circuit arrangements, which is commonly rationalized by surface inhomogeneity[46-48]. While $\alpha$ is usually left as a fit parameter, there are microscopic morphology models to predict its value[49-51]. Another approach is the phenomenological modification of the conventional models by a fractional exponent, e.g. $\coth(i\tau\omega)^{\beta/2}/(i\tau\omega)^{\beta/2}$ for planar bounded diffusion where $0<\beta<1$ [1,52-55], which could be attributed to surface roughness[41,56-58], hierarchical structures in porous electrodes[59], anomalous diffusion in disordered materials[57,60,61], or anisotropic diffusion in battery particles[32].

In many nanostructured materials, non-ideal diffusion impedance is also attributable to the inherent geometrical randomness, such as particle size distribution in batteries, pore size distribution in capacitors, tortuosity distribution in membranes and porous electrodes, and inhomogeneous boundary layer thickness in flow batteries. Such spatial heterogeneity naturally introduces distribution of diffusion times, corresponding to the set of internal paths lengths. Although this concept has been discussed in different contexts, including batteries[27,32,48,62-64], capacitors[59,65,66], fuel cells[67], and flow batteries[40], a general mathematical framework has not yet been developed to analyze experimental data. One approach is to couple several finite diffusion elements in parallel, as a crude approximation of the true heterogeneity[48,63]. Another approach is to assume a certain, continuous probability distribution function, sometimes based on supplemental observations such as electron microscope images[27,32,62,64-66]. Both approaches, however, require *a priori* knowledge about the nanostructure to properly choose the number of diffusion elements or the functional form of the distribution, which is usually not available.

In this Letter, we propose the theory of diffusion impedance for random heterogeneous materials based on a distribution of diffusion times (DDT). The innumerable diffusion paths in a nanostructure generally have different diffusion times, which often contribute independently in parallel to the collective diffusion impedance. Then, the generalized diffusion impedance is given by:

$$z_{GD}^{-1}(\omega) = \int_0^\infty P(\tau) z_D^{-1}(\omega, \tau) d\tau, \tag{1}$$

where $P(\tau)$ is the DDT, and $z_D$ is the finite-length diffusion model that represents the individual diffusion paths, as shown in TABLE I for a set of representative geometries and boundary conditions. We present a general method to solve the inverse problem for the DDT from experimental data that does not assume *a priori* knowledge on the configurational randomness, and consider three representative cases: a supercapacitor, a Li-ion battery, and a flow battery.

TABLE I. $z_D$ for different boundary conditions and symmetries. If a charge transfer reaction is taking place simultaneously, their Gerischer-type derivatives can be used instead.

| Boundary condition | Blocking | | | Transmissive |
|---|---|---|---|---|
| Symmetry | Planar | Cylindrical | Spherical | Planar |
| $z_D(\omega, \tau)$ | $\dfrac{\coth(\sqrt{i\omega\tau})}{\sqrt{i\omega\tau}}$ | $\dfrac{I_0(\sqrt{i\omega\tau})}{\sqrt{i\omega\tau} I_1(\sqrt{i\omega\tau})}$ | $\dfrac{\tanh(\sqrt{i\omega\tau})}{\sqrt{i\omega\tau} - \tanh(\sqrt{i\omega\tau})}$ | $\dfrac{\tanh(\sqrt{i\omega\tau})}{\sqrt{i\omega\tau}}$ |

The DDT framework generalizes the traditional interpretation of impedance spectra in terms of a distribution of relaxation times (DRT) for a linear superposition of the parallel RC circuit elements $(1 + i\tau\omega)^{-1}$ [68-70]. Recently, DRT analysis has been increasingly applied to electrochemical energy systems[54,71-74], although it is only intended to represent high-frequency interfacial charging and Faradaic reactions. Low-frequency diffusion impedance, which contains geometrical information about transport pathways, is mathematically and physically distinct and cannot be meaningfully represented by a DRT. Instead, the appropriate DDT must be defined.

The geometrical interpretation of the DDT, where each diffusion time, $\tau = l^2/D$, corresponds to a nanoscale path length $l$ traversed with diffusivity $D$, suggests a tantalizing possibility of "impedance imaging", by inverting the impedance spectrum to obtain $P(\tau)$. The inversion problem is a Fredholm integral equation of the first kind, which commonly appears in statistical thermodynamics[75], polymer rheology[76], medical imaging[77] and other fields. Upon a change of variables, such that $t = \log(\tau)$ and $u = -\log(\omega)$, Equation (1) can be written in a convolution form:

$$y(u) = \int_{-\infty}^{\infty} K(u-t) q(t) dt, \qquad (2)$$

where $y = z_{GD}^{-1}$ is the admittance of the nanostructure, $K = z_D^{-1}$ is the diffusion kernel determined by the diffusion conditions (TABLE I), and $q(t) = \tau P(\tau)$ is the unknown distribution function. Evaluating at a discrete set of $u_n$ for $n = 1, 2, ..., N$, and discretizing $t$ for a discrete set of $t_m$ for $m = 1, 2, ..., M$, Equation (2) can be approximated by $\mathbf{y} = \mathbf{KHq}$, where $\mathbf{y}$ and $\mathbf{q}$ are vectors such that $y_n = y(u_n)$ and $q_m = q(t_m)$; $\mathbf{K}$ is a kernel matrix such that $K_{n,m} = K(u_n - t_m)$; and $\mathbf{H}$ is a discretization matrix of the integral, for which we adopt the trapezoidal rule, although other discretization schemes may be used [78]. Given sufficient data, such that $N > M$, finding $\mathbf{q}$ may seem like a simple linear over-determined problem subject to an inequality constraint $\mathbf{q} \geq 0$. However, this class of inversion problems are known to be mathematically ill-posed, and a naïve least square regression does not provide a reliable solution.

Tikhonov regularization is a common methods to solve such an inversion problem[79,80], while others include the lasso regularization[81], the maximum entropy regularization[82,83], the Monte Carlo method[84], and the Fourier transform method followed by filtering[72,85]. Tikhonov regularization is a modified least square method where the loss function includes a penalty term that regulates one of the derivatives (Sobolev norm) of the solution. Here we opt to control the second derivative in order to smooth the fitting of irregular or noisy data. Given a vector of experimental data, $\mathbf{y}^\varepsilon$, the loss function has the following form:

$$\Phi(\mathbf{q}; \mathbf{y}^\varepsilon, \lambda) = \left\| \mathbf{W}(\mathbf{y}^\varepsilon - \mathbf{KHq}) \right\|_2^2 + \lambda \left\| \mathbf{D}_2 \mathbf{q} \right\|_2^2, \qquad (3)$$

where the first term is the conventional sum of residual squares, and the second term is the penalty term that imposes smoothness of the solution. Here, $\mathbf{W}$ is the diagonal weighting matrix, and $\mathbf{D}_2$ is the second order difference matrix that approximates $q''(t)$. $\lambda$ is the regularization parameter that determines the relative scale of the penalty term. Minimization of $\Phi(\mathbf{q})$ belongs to quadratic programming, and its standard formulation is presented in Supplemental Materials. Defining the intermediate solution function by $\mathbf{q}^\lambda(\lambda; \mathbf{y}^\varepsilon) = \arg\min \Phi(\mathbf{q}; \mathbf{y}^\varepsilon, \lambda)$ subject to $\mathbf{q} \geq 0$, it is largely affected by the value of $\lambda$; a small $\lambda$ results in overfitting or oscillation in $\mathbf{q}^\lambda$, whereas a large $\lambda$ results in over-smoothing.

To determine the optimal regularization parameter, $\hat{\lambda}$, we employed the real-imaginary cross-validation method, where we minimize the prediction errors of separate real and imaginary parts of $\mathbf{y}^\varepsilon$ with respect to the other part[81,86]:

$$\Psi(\lambda) = \left\| \mathbf{W}\left\{\operatorname{Re}[\mathbf{y}^\varepsilon] - \operatorname{Re}[\mathbf{KH}\mathbf{q}^\lambda(\lambda; \operatorname{Im}[\mathbf{y}^\varepsilon])]\right\}\right\|_2^2 + \left\| \mathbf{W}\left\{\operatorname{Im}[\mathbf{y}^\varepsilon] - \operatorname{Im}[\mathbf{KH}\mathbf{q}^\lambda(\lambda; \operatorname{Re}[\mathbf{y}^\varepsilon])]\right\}\right\|_2^2. \quad (4)$$

The nested minimization of $\Psi$ determines $\hat{\lambda}$, which then can be used to calculate the final solution, $\hat{\mathbf{q}} = \mathbf{q}^\lambda(\hat{\lambda}; \mathbf{y}^\varepsilon)$. This inversion method leads to an accurate estimation when a smooth, well-behaved solution is expected. Otherwise, the hierarchical regularization method can be employed[87].

We first illustrate the DDT method for artificial spectra generated from a known distribution with noise, using the inversion method to recover the distribution from the spectra. FIG. 1 shows two representative results where a normal distribution and a bimodal distribution are accurately determined from the spectra without *a priori* assumptions about the functional form. Further details of the simulation studies can be found in Supplemental Materials. The accuracy and resolution of the method depend on the noise level, as well as the completeness of the data. As shown in the bimodal example (FIG. 1 (d)), two distinct peaks can be resolved as long as $\Delta \bar{t} \Delta u = 1$ for $t_a \leq u_n \leq t_b$, where $\Delta \bar{t}$ is the separation of the peak times, $\bar{t}_a$ and $\bar{t}_b$, and $\Delta u$ is the sampling period among $u_n$. This uncertainty principle determines the resolution limit for impedance imaging, *e.g.* for a battery electrode with a mixture of active materials or a porous electrode with inhomogeneous local nanostructure. Even applying to their aging behavior, it could be possible to separately track the degradation of each part of the nanostructures. This, on the other hand, would be nearly impossible to discern in the impedance spectra, prior the inverse transform.

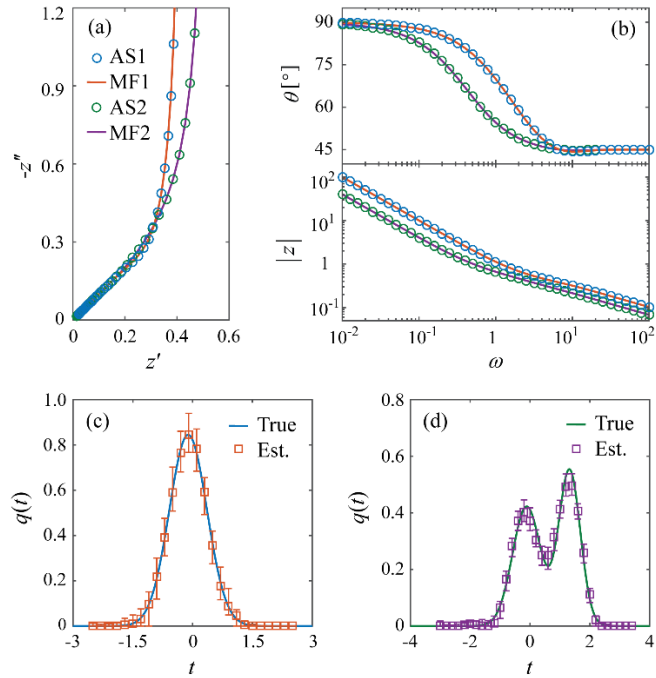

FIG. 1. (a) Artificial spectra (AS1 and AS2) and their model fits (MF1 and MF2) presented on the complex plane, and (b) their phase angles and magnitudes plotted against frequency. (c) The true normal distribution used in generating AS1 and its estimation by inversion of AS1, and (d) the true bimodal distribution used in generating AS2 and its estimation by inversion of AS2.

For the first physical example, we apply our method to determine the DDT for a vertically aligned carbon nanotube (CNT) supercapacitor from the experiments of Mutha *et al.* [88]. Considering a vertical unit space surrounded by the CNTs, conductive charging of the double layer along the CNT sidewalls can be described by the planar bounded Warburg kernel. The charging time is determined by the length of the tortuous CNTs and the cross-sectional area of the unit space, and their spatial variations lead to a DDT for the charging time. In FIG. 2 (a) and (b), the DDT model accurately fits spectra that deviate slightly from the ideal bounded Warburg behavior, and the underlying distribution is extracted by the inversion method. We can also see how the distribution changes by experimental variables, such as volume fraction ($V_f$) of CNTs as shown in FIG.2 (c). With increasing $V_f$, up to 15%, the primary distribution shifts to larger values in $t$. The most probable charging time, $\bar{\tau}$, shows a unit slope with respect to $V_f$ in a log-log plot (FIG. 2 (d)), which is predicted theoretically[88]. At a higher volume fraction, 26%, however, it shifts back to lower $t$, which is probably due to CNT bundling that renders the inner sidewalls inaccessible. Secondary peaks are observed in intermediate volume fractions from 2 – 10% as well as at 26%, which is associated with the long tails when the distributions are mapped onto the interspacing length, $\Gamma$, in FIG.2 (e). Such observation was not possible in the previous study assuming a normal distribution[88]. The distributions in $\Gamma$ obtained from the impedance inversion are consistent with a stochastic simulation[89-91].

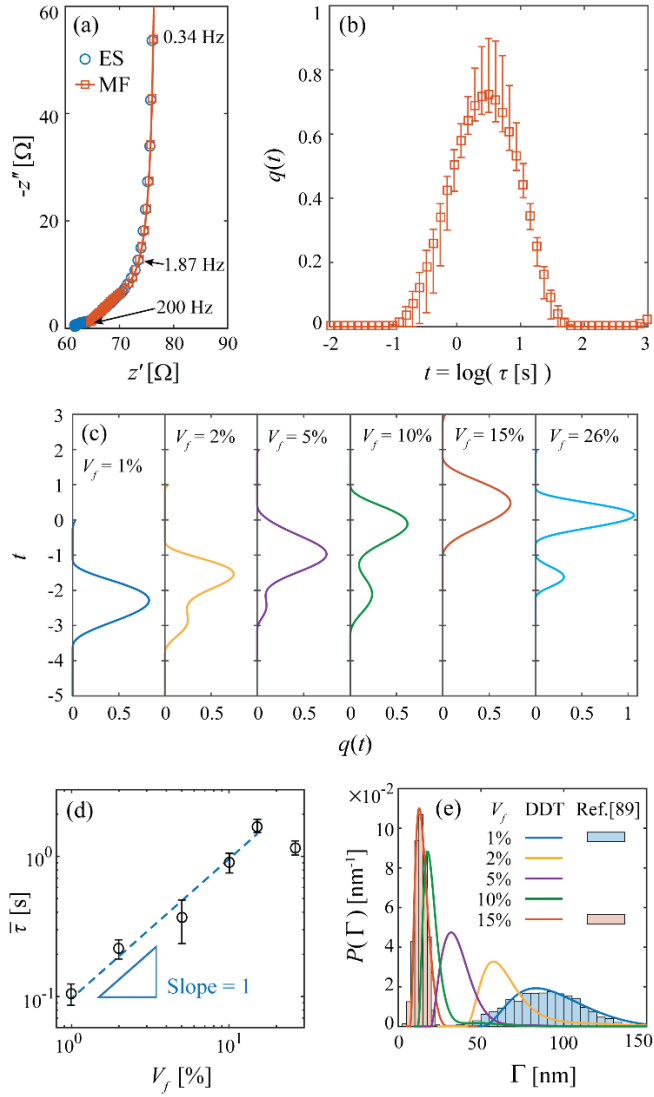

FIG. 2. (a) Experimental spectra (ES) of a CNT electrode and its model fit (MF) at $V_f$ = 15%. (b) The corresponding distribution of charging times estimated by the DDT model. Spectra at other $V_f$ and their inversion results are provided in Supplemental Materials. (c) Change in the distribution over a range of $V_f$, and (d) shift in the most probable time constant, $\bar{\tau}$, with respect to $V_f$. (e) Distributions in interspacing length, $\Gamma$, obtained by the DDT model and by a stochastic simulation[90].

For our second example, we perform a DDT analysis of impedance spectra for a silicon nanowire (SiNW) Li-ion battery anode from the work of Chan *et al.*[12] and Ruffo *et al.*[44]. Here, Li$^+$ diffuses radially from the side surface to the center of the nanowires, and the diffusion in individual nanowires is modeled by the cylindrical bounded diffusion kernel[27]. In FIG.3 (a) and (b), the DDT model accurately captures its inclined diffusion impedance, and the underlying DDT is determined by the inversion method. In FIG.3 (c) as the nanowires are lithiated, the DDT spreads wider, and then narrows back reversively after subsequent delithiation in FIG.3 (d). In FIG.3 (e), the DDTs are converted to nanowire radius ($r$) distributions, which show a consistent trend within the experimental observations. The Li$^+$ diffusivity can be estimated by matching either the mean ($\mu$) or the standard deviation ($\sigma$) of the radius distribution. As shown in FIG.3 (f), the obtained diffusivity varies between $1-5\times10^{11}$ cm$^2$/s, depending on the concentration, consistent with the results of Dimov *et al.*[92]. At high concentrations, both approaches result in impressively proximate estimations. On the other hand, the same spectra could result in overestimation if interpreted by a primitive model[44].

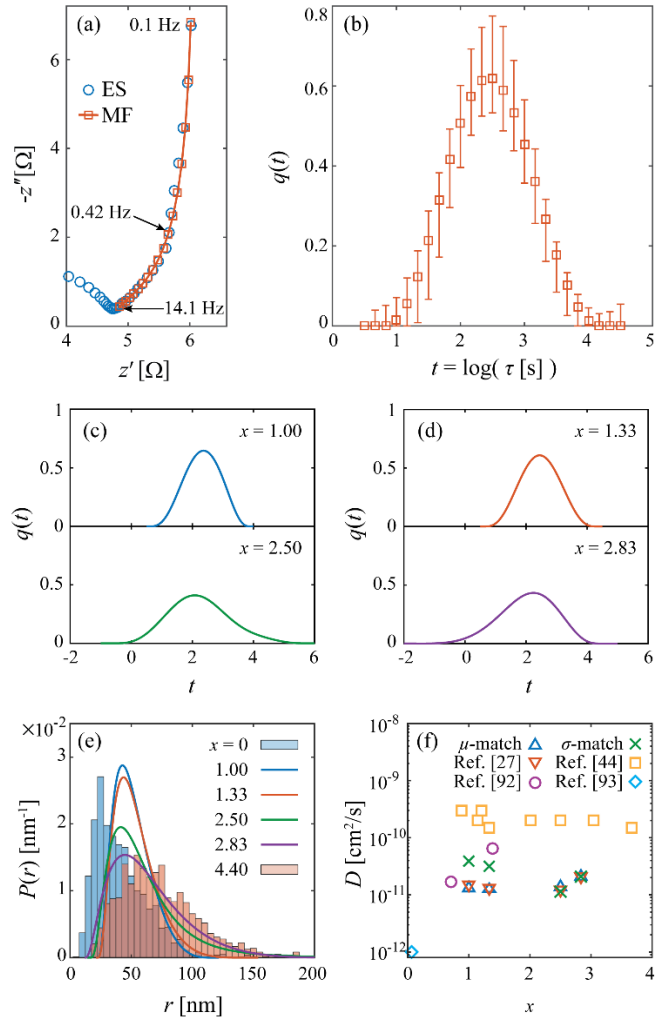

FIG. 3. (a) Experimental spectra (ES) of a SiNW electrode and its model fit (MF) at $x = 1.33$, where $x$ is the stoichiometric concentration of lithium in $Li_xSi$. (b) The corresponding DDT estimated by the inversion method. Spectra at other $x$ and their inversion results are provided in Supplemental Materials. (c) DDTs at a low and a high $x$ during lithiation, and (d) during subsequent delithiation. (e) Radius distributions obtained by the DDT model (curves) and the SEM image analysis (bars)[12]. (f) Estimated diffusivity and comparison to other studies[27,44,92,93].

Our final example, shown in FIG. 4, is a vanadium flow battery of Liu *et al.*[94], which illustrates DDT analysis for transmissive diffusion. As the electrolyte flows through the porous carbon electrode, boundary layer develops on the microscopic internal surface. The Nernst diffusion layer model leads to the transmissive diffusion kernel in TABLE I. The boundary layer has spatial variation in its thickness due to the variation in local velocity and pore configuration, which leads to a DDT. In FIG.4 (a), the model shows an excellent agreement with its diffusion impedance, even though the arc is significantly suppressed compared to the ideal finite-length Warburg behavior. The inverted DDT, shown in FIG.4 (b), reveals the dispersion of accessible paths for mass transport in a random porous medium, in addition to the mean. A more detailed study with microstructural characterization will follow[95]. Besides the examples, this approach is generally applicable to other electrochemical systems.

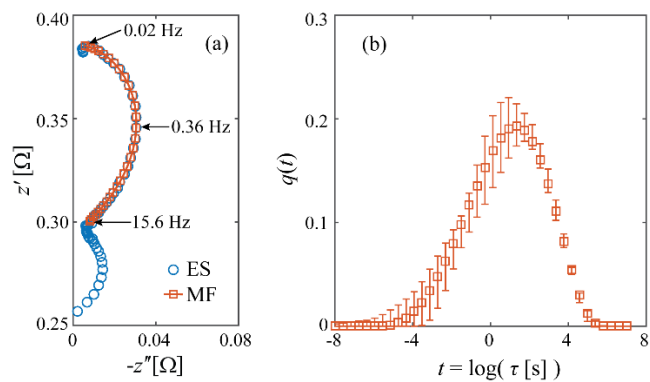

FIG. 4. (a) Experimental spectra (ES) of a vanadium flow battery and its model fit (MF). (b) The corresponding DDT estimated by the inversion method.

In conclusion, we have developed a mathematical framework to determine the DDT from electrochemical impedance spectra and demonstrated the possibility of impedance imaging for different types of nanostructured electrodes. The method is not limited to purely diffusive processes, but could be extended for reaction-diffusion phenomena in heterogeneous materials. For example, the model with a Gerischer-type kernel could be applied to impedance spectra for solid oxide fuel cells, in order to more accurately extract the surface diffusivity and adsorption rate constant for oxygen electrocatalysis by accounting for the observed statistical variations in the functional layer microstructure[34]. Our numerical inversion method could also be applied to hybrid DRT and DDT models, to simultaneously study the dispersion of low-frequency bulk diffusion and high-frequency interfacial charge and reactions.

ACKNOWLEDGEMENTS

This work was supported by the Legatum fellowship and the Kwanjeong fellowship (JS) and a Prof. Amar G. Bose Research Grant (MZB). The authors acknowledge Prof. Yi Cui at Stanford University, Heena K. Mutha, John L. Barton, and Prof. Fikile R. Brushett at MIT for sharing their experimental data and Itai Y. Stein for sharing his simulation results.

Supplemental Materials to

Electrochemical Impedance Imaging via the Distribution of Diffusion Times

Juhyun Song[1], and Martin Z. Bazant[1,2,*]

[1]*Department of Chemical Engineering and* [2]*Department of Mathematics,*

*Massachusetts Institute of Technology, Cambridge, Massachusetts 02139, United States*

1. Quadratic Programming Formulation

Minimization of $\Phi(\mathbf{q})$ in Eq. (3) can be reformulated into the standard form of quadratic programming[1] that only contains real quantities. Beginning by separating the real and the imaginary parts of the residual squares,

$$\Phi(\mathbf{q}) = \left\| \mathbf{W}\,\text{Re}(\mathbf{y}^\varepsilon - \mathbf{KHq}) \right\|_2^2 + \left\| \mathbf{W}\,\text{Im}(\mathbf{y}^\varepsilon - \mathbf{KHq}) \right\|_2^2 + \lambda \left\| \mathbf{D}_2 \mathbf{q} \right\|_2^2. \tag{S1.1}$$

Expanding the norm squares:

$$\begin{aligned}\Phi(\mathbf{q}) =\; &\mathbf{q}^\text{T} \left\{ \mathbf{H}^\text{T} \text{Re}[\mathbf{K}]^\text{T} \mathbf{W}^\text{T} \mathbf{W} \text{Re}[\mathbf{K}] \mathbf{H} + \mathbf{H}^\text{T} \text{Im}[\mathbf{K}]^\text{T} \mathbf{W}^\text{T} \mathbf{W} \text{Im}[\mathbf{K}] \mathbf{H} + \lambda \mathbf{D}_2^\text{T} \mathbf{D}_2 \right\} \mathbf{q} \\ &- 2\left\{ \text{Re}[\mathbf{y}]^\text{T} \mathbf{W}^\text{T} \mathbf{W} \text{Re}[\mathbf{K}] \mathbf{H} + \text{Im}[\mathbf{y}]^\text{T} \mathbf{W}^\text{T} \mathbf{W} \text{Im}[\mathbf{K}] \mathbf{H} \right\} \mathbf{q} \\ &+ \text{Re}[\mathbf{y}]^\text{T} \mathbf{W}^\text{T} \mathbf{W} \text{Re}[\mathbf{y}] + \text{Im}[\mathbf{y}]^\text{T} \mathbf{W}^\text{T} \mathbf{W} \text{Im}[\mathbf{y}].\end{aligned} \tag{S1.2}$$

Defining a new equivalent loss function, $2\Gamma(\mathbf{q}) = \Phi(\mathbf{q}) - \text{Re}[\mathbf{y}]^\text{T} \mathbf{W}^\text{T} \mathbf{W} \text{Re}[\mathbf{y}] - \text{Im}[\mathbf{y}]^\text{T} \mathbf{W}^\text{T} \mathbf{W} \text{Im}[\mathbf{y}]$, it now has the standard form of quadratic programing:

$$\Gamma(\mathbf{q}) = \frac{1}{2} \mathbf{q}^\text{T} \mathbf{A} \mathbf{q} + \mathbf{B}^\text{T} \mathbf{q}, \quad \text{such that } \mathbf{q} \geq 0, \tag{S1.3}$$

where

$$\begin{aligned}\mathbf{A} &= \mathbf{H}^\text{T} \text{Re}[\mathbf{K}]^\text{T} \mathbf{W}^\text{T} \mathbf{W} \text{Re}[\mathbf{K}] \mathbf{H} + \mathbf{H}^\text{T} \text{Im}[\mathbf{K}]^\text{T} \mathbf{W}^\text{T} \mathbf{W} \text{Im}[\mathbf{K}] \mathbf{H} + \lambda \mathbf{D}_2^\text{T} \mathbf{D}_2, \\ \mathbf{B} &= -\mathbf{H}^\text{T} \text{Re}[\mathbf{K}]^\text{T} \mathbf{W}^\text{T} \mathbf{W} \text{Re}[\mathbf{y}] - \mathbf{H}^\text{T} \text{Im}[\mathbf{K}]^\text{T} \mathbf{W}^\text{T} \mathbf{W} \text{Im}[\mathbf{y}].\end{aligned} \tag{S1.4}$$

The standard form can then be easily fed to commercial optimization tools.

2. Specification of the Simulation Study

The spectra in the simulation study (FIG. 1) are generated from known distribution functions with noise, to demonstrate the inversion method. The true distribution for the first artificial spectra (AS1) is a lognormal distribution in $\tau$ with a mean of 1.0 and a standard deviation of 0.5; in $t$-space, it is a normal distribution function with a mean of -0.11 and a standard deviation of 0.47, as shown in FIG. 1 (c). On the other hand, the true distribution for the second artificial spectra (AS2) is a bimodal distribution that combines two lognormal distributions in $\tau$ with respective means of 1.0 and 4.0, standard deviations of 0.5 and 1.5, and equal weights; the corresponding bimodal distribution in $t$-space is shown in FIG. 1 (d). The DDT model in Equation (2) is then employed with the planar bounded Warburg kernel (TABLE I) to generate the model spectra from the true distributions. The model is evaluated at logarithmically spaced frequencies for 20 points per decade (ppd) between $\omega=10^{-3}$ and $10^3$. Uncorrelated complex Gaussian noise is then introduced to obtain the artificial spectra, which has a zero mean and a relative standard deviation of 0.01 %. The resulting artificial spectra (AS1 and AS2) are plotted in FIG. 1 (a) and (b). Then the inversion method in the main text is employed to estimate the distributions only from the artificial spectra without *a priori* knowledge about the true distributions. The estimated distributions are compared to the respective true distributions in FIG. 1 (c) and (d). The mean absolute errors are 0.0016 for the normal distribution and 0.0032 for the bimodal distribution, respectively.

## 3. Inversion of Impedance Spectra of a CNT Electrode

In FIG. S1, impedance spectra of a CNT electrode at different volume fractions ($V_f$) are inverted by the model to estimate the distribution of charging times. The estimated distributions are then used in the analysis presented in FIG. 2 in the main text.

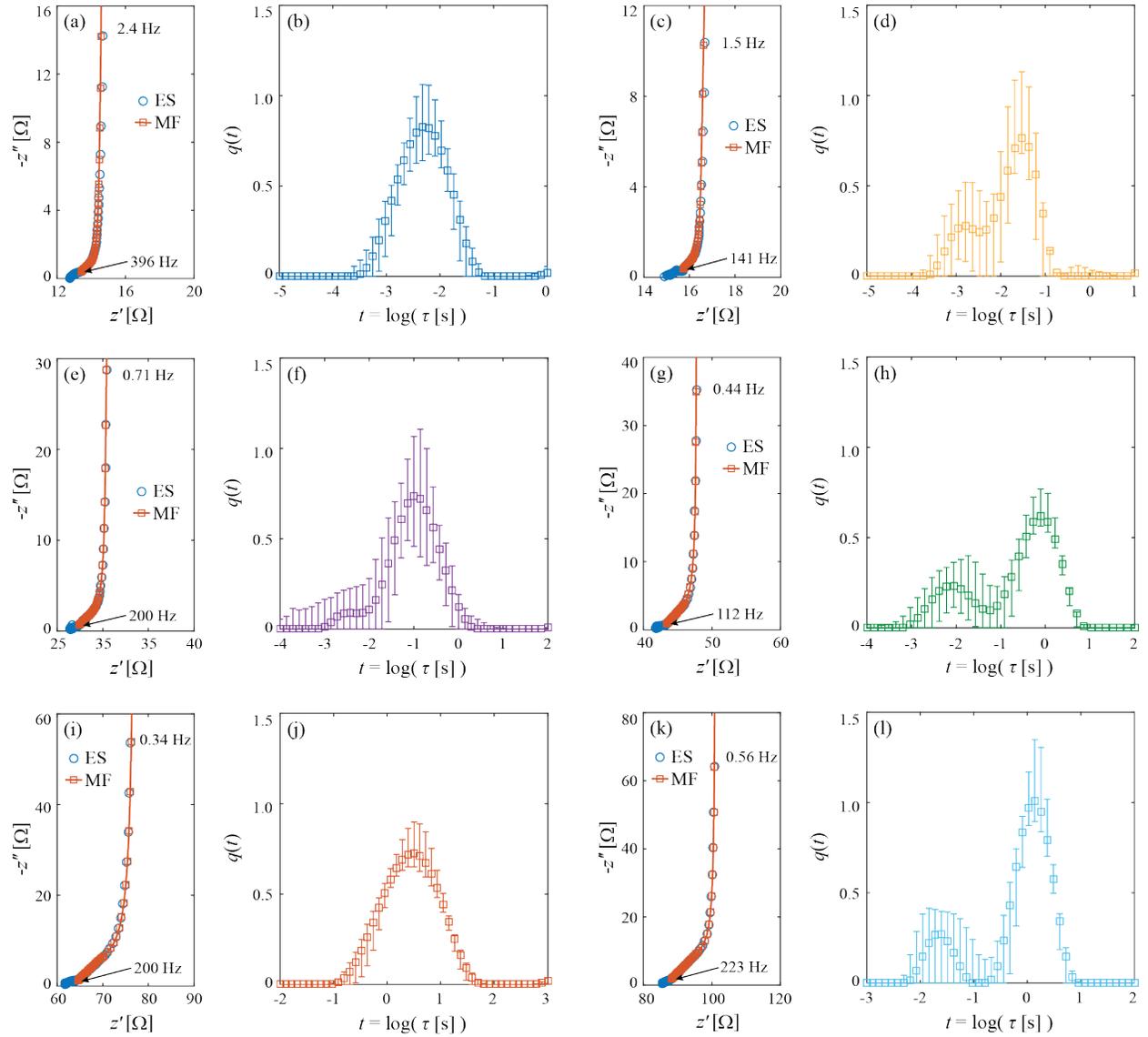

FIG. S1. Experimental spectra (ES) of a CNT electrode[2], their model fits (MF), and the corresponding distributions of charging times obtained by the DDT model: (a), (b) $V_f$ = 1%, (c), (d) $V_f$ = 2%, (e), (f) $V_f$ = 5%, (g), (h) $V_f$ = 10%, (i), (j) $V_f$ = 15%, and (k), (l) $V_f$ = 26%.

4. Inversion of Impedance Spectra of a SiNW Electrode

In FIG. S2, impedance spectra of a CNT electrode at different lithium concentrations are inverted by the model to estimate the distribution of diffusion time. The estimated distributions are then used in the analysis presented in FIG. 3 in the main text.

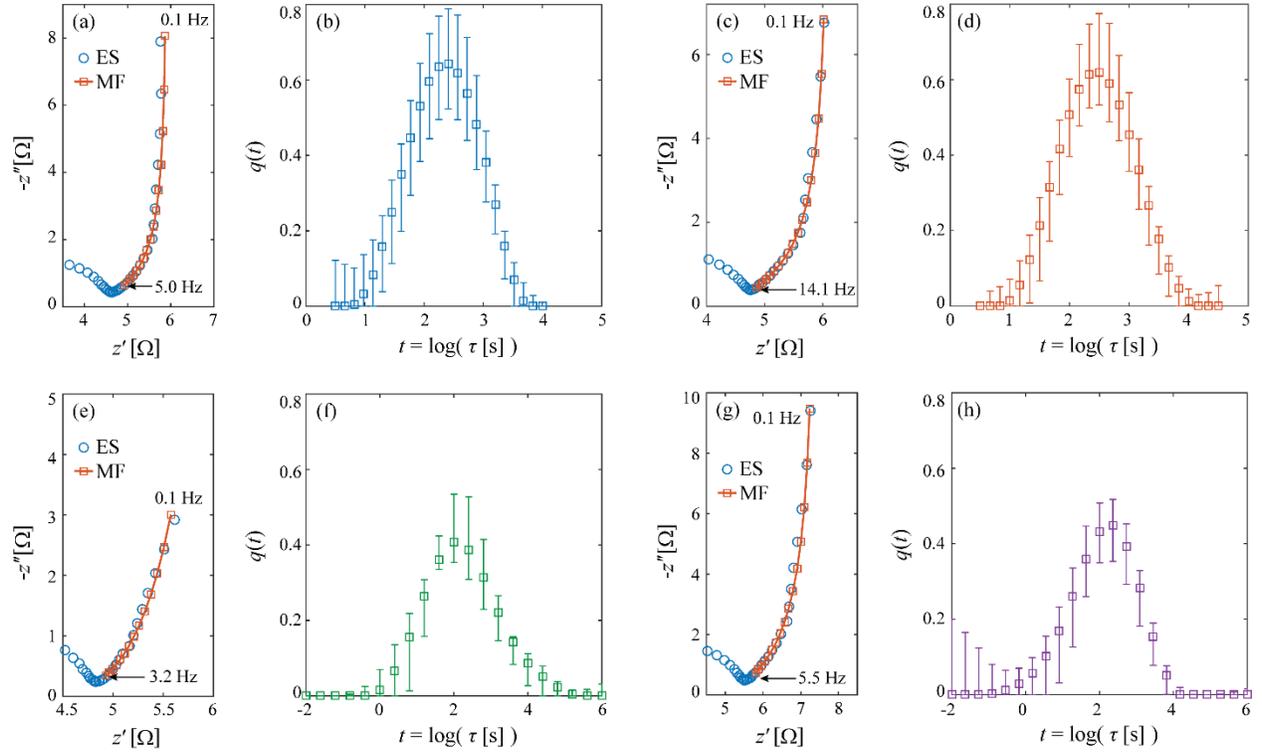

FIG. S2. Experimental spectra (ES) of a SiNW electrode[3], their model fits (MF), and the corresponding DDTs obtained by the inversion method: (a), (b) $x=1.00$; (c), (d) $x=1.33$; (e), (f) $x=2.50$; (g), (h) $x=2.83$, where $x$ is the stoichiometric concentration of lithium in $Li_xSi$.